\documentstyle[prl,aps,multicol]{revtex}                                       

\begin{document}

\title{Strict Bounds on Franson Inequality}

\author{ Lev Vaidman and Ori Belkind} 

\address{School of Physics and Astronomy, 
Raymond and Beverly Sackler Faculty of Exact Sciences, \\
Tel-Aviv University, Tel-Aviv 69978, Israel.} 

\date{}

\maketitle

\begin{abstract}
An inequality, recently proposed by Franson [Phys.  Rev. {\bf  A 54},
3808 (1996)] is analyzed and improved. The inequality 
connects  the change of the  expectation value of an observable
with the uncertainty of this observable. A strict bound on the ratio
between these two quantities 
 is obtained.
\end{abstract}


\begin{multicols}{2}

In a recent paper Franson\cite{Fr} developed a quantitative expression
which characterizes the fact that  a quantum observable cannot change
its expectation value without having some  quantum uncertainty. The simplest
explanation of this property is that in order for a quantum system to go
from one eigenvalue of an observable to another it must evolve through
a superposition of the eigenstates which is characterized by nonzero
uncertainty of the observable. Franson proposed to consider the change of
an observable $Q$ after it has  been measured such that the initial
state of the system at time $t=0$ is an eigenstate of the
observable. Then, at any moment of time $t$ in the period  $\tau_1$,
 \begin{equation}
   \label{tau}
\tau_1   = {\hbar \over {\sqrt 2 \Delta E}} ,
 \end{equation}
 where $\Delta E$ is the
 energy uncertainty of the system, the following inequality is  fulfilled:
\begin{equation}
  \label{fr-in}
  |\delta \langle Q \rangle| \leq \Delta Q .
\end{equation}
Here, $\Delta Q = \sqrt{  \langle \Psi(t) |Q^2|\Psi(t)\rangle - \langle \Psi(t) |Q|\Psi(t)\rangle^2}$ is the uncertainty at time $t$ and  $\delta \langle Q
\rangle = \langle \Psi(t) |Q|\Psi(t)\rangle - \langle \Psi(0)
|Q|\Psi(0)\rangle$ is the change in the expectation value.

If an inequality expresses a basic physical law, there must be cases
where it is saturated; if not, it must be replaced by a stricter
inequality which does saturate. The Franson inequality as it is
presented in his paper cannot reach the equality (except at $t=0$).
The purpose of this paper is to improve Franson's bound such that the revised
inequality cannot be replaced by any stricter inequality.

There are two elements which we  improve in the Franson
inequality. First, we replace his approximate calculations by exact
calculations and obtain, instead of (1), a larger period of time $\tau_2$
for which the inequality (2) is fulfilled. Then, for time $t=\tau_2$ the
inequality can be saturated. However, for any intermediate time, $t\in
(0,\tau_2)$, the inequality  cannot be saturated. In order to
correct this we introduce time  explicitly into the the inequality.

The Franson inequality is closely connected with a certain type of time-energy
inequality which constrains  the time of the evolution of a system to an
orthogonal state:
\begin{equation}
t\geq {{h}\over {4 \Delta E}}.
\end{equation}
 Although it has a similar form, it is conceptually
different from the Heisenberg uncertainty relations for position and
momentum.\footnote{Recently, Yu\cite{Yu} described an analog of this
  time-energy inequality for position and momentum.}
There are many derivations of
  this results. As far as we know the first derivation was given
  by Mandelstam and Tamm\cite{MT}, and a very simple  derivation can
  be found in
  Vaidman\cite{V}. 
We will apply this result for deriving  a strict bound on the Franson
inequality. A more general form of the time-energy inequality which  we
will need here is:
\begin {equation}
|\langle \Psi (t)|\Psi (0)\rangle| \geq \cos ({{\Delta E ~ t}\over
  \hbar}) ,
\end{equation}
which is valid for $t\in [0,{{\pi\hbar}\over {2\Delta E}}]$ where
$\Delta E$ is the energy uncertainty of the system.

Let us turn to  finding the optimal bound on the period of time for
which the Franson inequality (2) is fulfilled. This problem is equivalent to finding
the minimal (nonzero) time for which $ |\delta \langle Q \rangle| = \Delta Q $.
Without loss of generality we can assume that the eigenvalue of $Q$ at
the initial time is zero, i.e. $|\Psi(0)\rangle =~|Q=0\rangle\equiv |0\rangle$. At time
$t$ the state of the system  can be expressed as  
\begin{equation}
|\Psi(t)\rangle = \alpha|0\rangle + \beta |1\rangle ,
\end{equation}
where $|1\rangle$ is orthogonal to $|0\rangle$  and $\beta$ is real
 and positive. Then,
\begin{eqnarray}
\delta \langle Q\rangle &=&\beta^2 \langle 1 |Q| 1 \rangle ,\\
\Delta Q &=& \sqrt {\beta^2 \langle 1 |Q^2| 1 \rangle-\beta^4  
\langle 1 |Q| 1 \rangle^2} .
\end{eqnarray} From $|\delta \langle Q\rangle| = \Delta Q $ follows:
\begin{equation}
\beta^2 ={{1}\over{2}} {{\langle 1 |Q^2|1 \rangle }\over
{\langle1 |Q|1 \rangle^2 }}
\end{equation}
Since $\langle1 |Q|1 \rangle^2$ is always
  smaller or equal to $\langle1 |Q^2|1 \rangle$,
    the minimal value for $\beta$ is $\sqrt {1\over 2}$.
For this value of $\beta$ we obtain:
\begin{equation}
  \label{calc}
  |\langle \Psi (t)|\Psi (0)\rangle| = |\alpha| = \sqrt{1 - \beta^2} =
  \cos {{\pi}\over{4}} .
\end{equation} From the time-energy relation (4)  it follows that the minimal time (which
corrects Franson's bound (1)) is:
 \begin{equation}
\tau_2 = {{\pi\hbar}\over{4\Delta E}}.
\end {equation}
Since $1/{\sqrt 2} < \pi/4$, the time limit we found is indeed
larger than that of Franson (Eq. \ref{tau}). Note, that this is half
the minimal time of an evolution to an orthogonal state.

Consider an example of a spin-$1\over 2$ particle precessing in
magnetic field. Let us assume  that the initial state is
$|{\uparrow}\rangle$, the ``up'' state in the $z$ direction, and that the magnetic field  is in the $y$ direction.
Then, the time evolution is: 
\begin{equation}
  \label{t-e}
  |\Psi(t)\rangle = \cos ({{\Delta E ~ t}\over \hbar})
  ~|{\uparrow}\rangle + \sin ({{\Delta E ~ t}\over \hbar})
  ~|{\downarrow}\rangle .
\end{equation}
Therefore,
\begin{eqnarray}
\delta \langle \sigma_z\rangle &=& \cos ({{2\Delta E ~ t}\over \hbar}) -
1 ,
\\
\Delta \sigma_z &=&  \sin ({{2\Delta E ~ t}\over \hbar}) .
\end{eqnarray}
Thus we see that $|\delta \langle \sigma_z\rangle| \leq \Delta
\sigma_z$    for $t\leq \tau_2$. The equality is reached  for   $t
=\tau_2$. 

The same bound is obtained for  any other two-level system or even
for  any
system with higher-dimensional Hilbert space  which spans only a 
two-dimensional Hilbert space during the evolution. Introducing any additional state will
invariably increase the minimal time for reaching the equality $|\delta \langle Q\rangle| = \Delta
Q$.

It is   interesting to consider the  connection between $\delta \langle
Q\rangle $ and $\Delta Q$  when  we do not impose the initial condition
$\Delta Q =0$. We  ask what is the minimal period of  time during
 which the change in the expectation value reaches the value of the
 maximal uncertainty during this period of time.

Consider  our example 
of a spin-$1\over 2$ particle. A simple analysis shows that  the
minimal time to reach the equality  $|\delta \langle Q\rangle| = \Delta
Q$ is obtained for  the evolution between
the  states
\begin{eqnarray}
  \label{i-f}
  |\Psi_{in}\rangle &=& \cos  ({\pi\over 6})
  |{\uparrow}\rangle + \sin ({\pi\over 6})  |{\downarrow}\rangle,\\
 |\Psi_f\rangle &=& \cos  ({\pi\over 3})
  |{\uparrow}\rangle + \sin ({\pi\over 3})  |{\downarrow}\rangle.
\end{eqnarray}
This  time  is
\begin{equation}
  \label{tau3}
 \tau_3 = {{\pi\hbar}\over {6\Delta E}}. 
\end{equation}
In this case  $|\delta \langle \sigma_z\rangle| = \max ({\Delta
\sigma_z)} =1$ and the maximum uncertainty is obtained at the middle
point of the evolution. 

However, an example in which only two states are involved is not the
optimal one. A system of $N$ spin-$1\over 2$ particles precessing in a
magnetic field yields
a smaller time for reaching the equality   $|\delta \langle \sum_i (\sigma_z)_i
\rangle| = \Delta (\sum_i (\sigma_z)_i)$. The minimal time for this is: 
\begin{equation}
  \label{tau4}
 \tau_4 = {\sqrt N} \arcsin({{1\over{2\sqrt N}}}){{\hbar}\over{\Delta E}}. 
\end{equation}
At the limit of $N\rightarrow \infty$ we reach the minimal time
\begin{equation}
  \label{tau5}
 \tau_5 = {{\hbar}\over{2\Delta E}} . 
\end{equation}
The same minimal time is obtained in the most natural example of
comparison between  $\delta \langle
Q\rangle $ and $\Delta Q$. Consider  a free particle  of mass $m$ in one dimension
in a (minimal
uncertainty) Gaussian wave packet which moves with velocity $v$. The
velocity 
is high  enough such that we
can neglect the spread  of the wave-packet.    
Since $E =p^2/2m$, for large velocity  we obtain 
\begin{equation}
  \label{DE}
\Delta E \sim {{\langle p\rangle \Delta p} \over m} .
 \end{equation}
 Then, taking into account the Heisenberg relation for minimal
 uncertainty, $\Delta x \Delta p = \hbar/2$, we find the time for 
which  $\delta \langle x\rangle = \Delta x$:
\begin{equation}
  \label{vel}
{\Delta x \over v} =  {\Delta x~ m \over \langle p\rangle} = {{\Delta x
    \Delta p}\over {\Delta E}} = {{\hbar}\over{2\Delta E}}.
\end{equation}

The period $\tau_5$ is a strict bound for  the minimal period of  time during
which the change in  the expectation value reaches the value of the
 maximal uncertainty. This result can be
 obtained immediately from the Heisenberg relation:
 \begin{equation}
   \label{heis}
 \Delta E \Delta Q \geq {1\over 2} |\langle [H,Q]\rangle | = {\hbar
   \over 2} | {{d\langle  Q\rangle}\over {dt}} | . 
 \end{equation}
If $\Delta Q$ is essentially constant during the process, we find 
that the equality $ | \delta \langle Q\rangle| = \Delta Q$ is reached
during the time $\tau_5$.

The strict bound $\tau_5$ and the bounds $\tau_3$ and $\tau_4$ are
bounds for a different problem from the one  Franson proposed. The novelty
of his inequality is in  considering  an evolution starting from an
eigenstate of an observable. For such a problem
only the Franson bound $\tau_1$ and the improved bound $\tau_2$ are
relevant. The bound $\tau_2$ is absolute in the sense that we cannot
replace it by a smaller value and the inequality (2) would still be
fulfilled for the whole period $t \in [0, \tau]$. However, even with
this exact bound the inequality cannot be considered as a fundamental
one because it cannot be saturated for any time except $t= 0$ and
$t =\tau_2$. 
 In order to find a basic inequality let us return  to Eqs. (6) and (7),
 but now  we will not limit ourselves to the equality $|\delta \langle
 Q\rangle| = \Delta Q $. By dividing the equations we obtain:
 \begin{equation}
   \label{div}
{{|\delta \langle Q\rangle|}\over  \Delta Q} =
 {1\over {\sqrt {{{\langle 1 |Q^2|1 \rangle }\over
{\beta ^2 \langle1 |Q|1 \rangle^2 }} -1}}} .
\end{equation} From  the time-energy relation (4) we obtain
\begin{equation}
\beta = \sqrt {1-|\langle \Psi(0)|\Psi(t) \rangle |^2} \leq \sqrt {1- 
  \cos^2 ({{\Delta E ~ t}\over \hbar})}
 =\sin ({{\Delta E ~ t}\over \hbar}) .  
\end{equation}
Now, taking again into account that $\langle1 |Q|1
  \rangle^2
\leq \langle1 |Q^2|1 \rangle$ we obtain the new inequality:
\begin{equation}
 {{|\delta \langle Q\rangle|}\over  \Delta Q} \leq \tan({{\Delta
     E}\over \hbar} t) .
\end{equation}
This inequality is valid for $t\in (0,{{\pi\hbar}\over {2\Delta
    E}}]$. The inequality (24) is a basic law since it cannot be
replaced by a stricter inequality. Indeed, the example of a
spin-$1\over 2$ particle precessing in a 
magnetic field saturates the inequality: it becomes an equality for the
whole period  $(0,{{\pi\hbar}\over {2\Delta
    E}}]$.

It is a pleasure to thank  Lior Goldenberg and Aharon Casher  for helpful discussions.  The research was
supported in part by grant 614/95 of the Israel Science
Foundation. Part of this work was completed during the 1997
Elsag-Bailey -- I.S.I. Foundation research meeting on quantum computation.

\end{multicols}

\end{document}